\documentclass[twocolumn]{aastex63}

\received{2019 October 16}
\revised{2019 December 11}
\accepted{2019 Deceber 17}
\published{2020 January 16}

\shortauthors{Rout et al.}

\graphicspath{{./}{figures/}}

\begin{document}

\title{A retrograde spin of the black hole in MAXI J1659--152}

\correspondingauthor{Sandeep K. Rout}
\email{skrout@prl.res.in, skrout92@gmail.com}

\author[0000-0001-7590-5099]{Sandeep K. Rout}
\affiliation{Physical Research Laboratory, Navarangpura, Ahmedabad 380009}
\affiliation{Indian Institute of Technology, Gandhinagar, Gujarat 382355}

\author[0000-0002-2050-0913]{Santosh Vadawale}
\affiliation{Physical Research Laboratory, Navarangpura, Ahmedabad 380009}

\author[0000-0003-2187-2708]{Mariano M\'endez}
\affiliation{Kapteyn Astronomical Institute, University of Groningen, P.O. Box 800, 9700, AV Groningen, The Netherlands}

\begin{abstract}

We present the results of spectral analysis of the galactic black-hole binary MAXI J1659$-$152 in the rising phase of the outburst that lasted for about 65 days starting on 2010 September 25. The presence of a broad Fe line, verified by Monte-Carlo simulations, and coverage of a wide energy band by utilizing the combined spectral capabilities of XMM-Newton/EPIC-pn and RXTE/PCA allowed us to use a combination of reflection spectroscopy and continuum fitting methods to estimate the spin of the black hole. We explored the entire parameter range allowed by the present uncertainties on black-hole mass, inclination, and distance as well as the accretion rate. We show that for about $95\%$ of parameter space and very reasonable upper limits on $\dot{M}$, the spin of the black hole has to be negative. This is the first clear detection of negative spin in a galactic black-hole binary. 

\end{abstract}

.

\section{Introduction} \label{sec:intro}

Stellar-mass black holes, harboring a low-mass star ($\sim 1 M_\odot$) in a binary system, spend most of their life in quiescence. They become visible in the X-ray sky only during violent episodes of outbursts triggered by an instability in the accretion disk \citep{frank02}. It is during these outbursts that many properties of the system, especially pertaining to the black hole and the accretion flow, are studied. A black hole can be characterized by two classical properties - mass, and angular momentum. A third property, namely electric charge, is assumed to be negligible in most astrophysical settings \citep{gurlebeck15}. While robust estimates of the black-hole mass is achieved by radial-velocity measurements of the secondary, accurate measurement of the spin is tricky, partly because of its dependence on the knowledge of various system parameters. Primarily, the spin of a black hole can be measured by two techniques - the continuum fitting method \citep[CF]{zhang97}, and Fe-line spectroscopy \citep{fabian89}. Both methods infer the value of spin indirectly by measuring the inner radius of the accretion disk, which is assumed to extend down to the innermost stable circular orbit (ISCO). In the CF method, the inner radius is estimated by fitting the thermal disk continuum with a general relativistic disk model  \citep{gierlinski01,mcclintock06,shafee06}. Geometrical parameters of the system like black-hole mass, distance, and inclination and the spectral hardening factor must be known a priori for the CF method to work. Reflection spectroscopy consists of modelling the spectrum originating from the reflection of the back-scattered coronal emission from the inner disk. Two important features of this spectrum are the fluorescent Fe-K$\alpha$ emission between 6.4 to 6.97 keV and a Compton hump peaking at around 30 keV. The red-ward extent of the line profile, that gets skewed by gravitational redshift, essentially gives the inner radius of the disk and hence the spin \citep{iwasawa96,miller02,miller04}, while the blue wing of the line essentially gives a measure of the inclination \citep{miller18}. There have also been attempts to constrain the black-hole spin using quasi periodic oscillations (QPO). The relativistic precesion model \citep[RPM;][]{stella98,ingram14} associates various QPO frequencies with the orbital and precesion frequencies of the accretion disk. \citet[][]{motta14a} and \citet{sramkova15} have applied this method to arrive at an estimate of the spin. However, these measurements remain few in number and enshrouded by uncertainty owing to there dependence on the models used. \citet[][]{dovciak08} showed that the polarisation angle and degree, expressed as a function of thermal energy, varies with spin and can be used as a method for spin determination, although it is yet to be applied.

\begin{figure*}[t]
    \centering
    \includegraphics[scale=0.6]{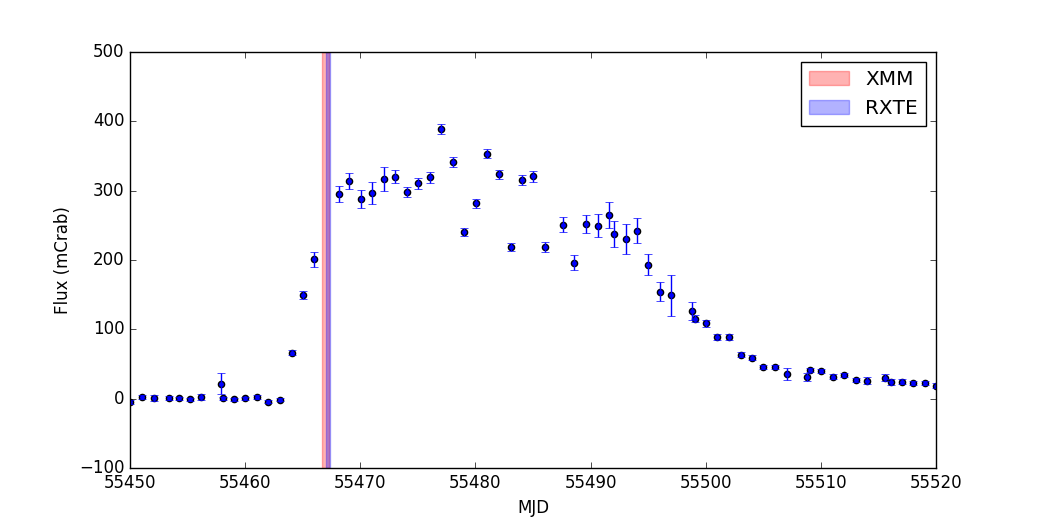}
    \caption{J1659 lightcurve in the 2-10 keV band with MAXI/GSC \citep{matsuoka09}. The colored vertical bars represent the epochs of the XMM-Newton and RXTE observations.}
    \label{fig:maxilc}
\end{figure*}

While the spin of a stellar mass black hole is reminiscent of the natal kick during a supernova explosion, the angular momentum of the accretion disk is determined by the binary orbit. Although, accretion would tend to align both the spin and the disk angular momentum through torques, that generally does not happen in LMXBs. This is because the mass required to be gained by the black hole to alter the spin significantly can not be supplied by a low-mass companion during the binary evolution \citep{king99,mcclintock06}. Thus, the spin is most likely natal and should be randomly distributed among the black-hole population. However, on observational grounds, most of the sources that have robust spin measurement shows positive spin \citep{mcclintock14,reynolds14}. It is only recently that a few systems with negative, or being consistent with negative, spin have come up \citep{morningstar14a,reis13,rao12}. Here, we present a case where the binary system MAXI J1659--152 hosts a negative spin black hole for almost the entire parameter space.

\section{Observation and Data Reduction} \label{sec:observation}

MAXI J1659$-$152 (henceforth termed as J$1659$) went into outburst on 2010 September 25, and was detected by Swift/BAT at 08:05 UTC on that day. Soon after the initial detection, Swift and RXTE were used to monitor the source continuously revealing many important properties of the source \citep{yamaoka12, kennea11, kalamkar11}. Three days into the outburst J$1659$ was observed with XMM-Newton with a single pointing of $\sim$ 50 ks exposure. The availability of simultaneous XMM-Newton and RXTE data enables us to try both the continuum fitting and reflection spectroscopy to measure the spin of the black hole.

The complete outburst lightcurve of J$1659$ is shown in figure \ref{fig:maxilc}. The source reached its maximum luminosity on MJD 55477 during a flare and the thermal peak on MJD 55489 \citep{kalamkar11}. J$1659$ was observed in timing mode with XMM-Newton \citep{jansen01} on September 27 at UTC 16:15:27. For our analysis, only the pn-CCD of the European Photon Imaging Camera was used \citep{struder01}. We used the recent version of SAS (17.0.0) and followed the instructions given in the data analysis threads \footnote{https://www.cosmos.esa.int/web/xmm-newton/sas-threads}. Standard procedure for pile-up correction was undertaken by excising central 5 rows of the PSF and comparing the grade ratios from the output of the SAS tool \texttt{epatplot}.

It was found that the background was contaminated by source counts because of its brightness and the relatively extended PSF of the EPIC-pn CCD. The flux difference between the background corrected and uncorrected spectrum was found to be $\sim 3.7\%$. It was verified using phenomenological models that the inclusion or exclusion of background did not have any significant impact on the model parameters. Hence, the analysis was carried out with the background extracted from the tail of the PSF. The data were grouped to have a minimum of 25 counts per bin to facilitate chi-square statistics and a systematic error of $1.5 \%$ was added. It is quite customary to encounter absorption features of instrumental origin in the EPIC-PN (timing mode) spectrum at $\sim$1.8 and $\sim$2.3 keV corresponding to Si-K and Au-M edges respectively \citep{wang19, papitto09}. To keep the model simple, we ignored the range of 1.5 - 2.5 keV from the spectrum instead of adding two absorption components which, we verified, would not have improved our results significantly.

J$1659$ was observed with RXTE \citep{swank06} on several occasions across the outburst. One observation with Obs. ID 95358-01-02-00 starting on 28 September 2010 at 00:58:24 partially overlapped with the XMM-Newton observation for a duration of $\sim$20 ks. Standard screening and filtering criteria were used to analyze the data of the Proportional Counter Array \citep[PCA,][]{jahoda06}. Only the top layer of PCU2 was used for the analysis. A bright model for the background as provided in the PCA Digest page\footnote{https://heasarc.gsfc.nasa.gov/docs/xte/pca\_news.html} was used to produce the background spectrum. The exposure of both the source and background spectra were corrected for dead time effects and a systematic error of $0.5 \%$ was added. Since the observation was during the rising phase of the outburst (figure \ref{fig:maxilc}) only the overlapping period of PN and PCA data, with an exposure of $\sim$20 ks, was used to avoid any spectral change. Upon using the full range of PCA, large residuals were observed in the 4 - 10 keV range. Such features are due to energy dependent cross-calibration uncertainty between PN and PCA and were previously reported by \citet[][]{kolehmainen14} and \citet{hiemstra11}, among others. Thus, we use the PCA in the 10-40 keV range. 

\section{Analysis and Results} \label{sec:results}

\begin{figure*}
    \centering
    \includegraphics[scale=0.8]{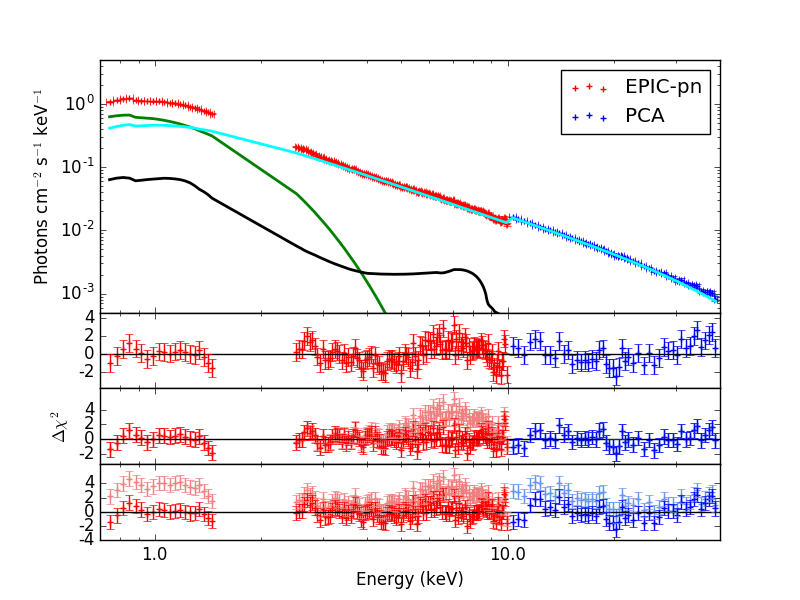}
    \caption{Top panel: The unfolded spectrum of J1659 along with the individual model components. The cyan, green, and black curves represent, respectively, the thermal-disk, powerlaw and reflection components (Model 3 below). Residuals for the following models from second to fourth panel: Model 1 - \texttt{const*phabs* (diskbb + nthComp)}; Model 2 - \texttt{const*phabs (gaussian + diskbb + nthComp)}; Model 3 - \texttt{const*phabs*(diskpn+nthComp+relxillCp)}. The lighter-shade residuals in the bottom two panels were obtained by fixing the normalization of \texttt{Gaussian} and \texttt{relxillCp} to zero respectively.}
    \label{fig:delchi}
\end{figure*}

\begin{figure*}
    \centering
    \includegraphics[scale=0.35]{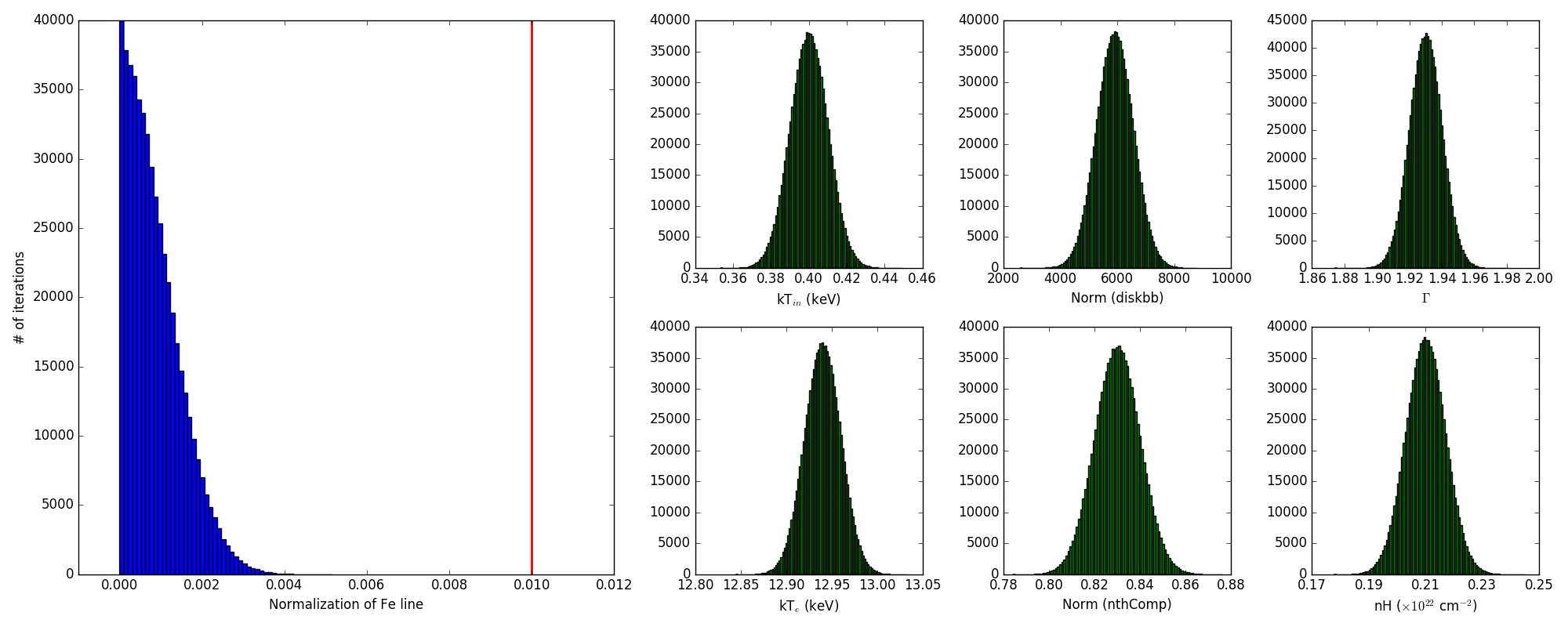}
    \caption{Results of the monte-carlo simulations to test the significance of the Fe emission line in J1659. The left panel shows a comparison of the norms obtained from simulations (blue histogram) to that obtained from the data (red line). The first bin (left most) in the histogram reaches up-to 500000, and has been clipped at 40000 for clarity. The six plots in the right show the distribution of the parameters that went into the simulation.}
    \label{fig:mcsim}
\end{figure*}

We fitted the data in \texttt{XSPEC}- v12.10.1 with a combination of models to describe the broad-band spectrum of J$1659$. The best fitting parameters for each of the models considered are listed in table \ref{table:bestfit_par} and the residuals of the fits are shown in figure \ref{fig:delchi}. To allow for the possible energy-independent cross-calibration uncertainties a multiplicative factor, \texttt{constant}, was added to the model. This parameter was frozen at 1 for EPIC-pn and left free to vary for PCA. The photo-electric absorption in the inter-stellar medium, was accounted by multiplying a \texttt{phabs} component to all the models. \\
We first fitted the Comptonisation model \texttt{nthComp} \citep{zdziarski96} but the fit was unacceptable with $\chi^2_\nu = 12.27$ for 186 degrees of freedom (dof). The fit was repeated with a blackbody-like model to account for the thermal (disk) component of the spectrum. Adding a \texttt{diskbb} model \citep{mitsuda84} gave a better fit than before, with $\chi^2$/d.o.f. $= 244.60/185$. However, the second panel of figure \ref{fig:delchi}) show positive residuals at $\sim 7$ keV, which are most likely due the Fe-K$\alpha$ emission. To incorporate this feature, a \texttt{Gaussian} component was added to the model that improved the fit significantly (see figure \ref{fig:delchi}, second panel) along with keeping most of the other parameters within 90\% confidence of the previous fit, and yielding a $\chi^2_\nu = 0.76$ for 182 d.o.f. The best-fit parameters for both these models are given in table \ref{table:bestfit_par}. The F-test probability for the \texttt{Gaussian} being present by chance was $3.93 \times 10^{-22}$. However, the F-test is not always appropriate for verifying the significance of line models \citep{protassov02}. Thus, a Monte-Carlo simulation was carried out for the same. In this regard, the best-fit continuum model, without the line, was used to simulate a series of $10^6$ spectra by incorporating the uncertainties in the continuum parameters from the previous fit. Then these spectra were fitted with a model including a \texttt{Gaussian} component with the line energy and width fixed to the respective best-fit values from Model 2 and its norm compared with the best fit norm. We never found a case in which the norm was equal or larger than the one in Model 2, hence we conclude that the probability for the spectrum to fit the line component by chance is less than $10^{-6}$. The results of this exercise are plotted in figure \ref{fig:mcsim}.   

A broad Fe line is a strong signature of reflection from regions close to the black hole, the broadening being essentially caused by gravitational redshift and Doppler effects \citep{fabian00}. This motivated us to use the state-of-the-art reflection code of the \texttt{relxill} family so as to constrain the spin of the black hole \citep{dauser14,garcia14}. The flavor that was opted, i.e., \texttt{relxillCp}, assumes a coronal geometry with a broken power law emissivity which was fixed to 3 for the entire disk. We replaced the thermal disk component, \texttt{diskbb}, with \texttt{diskpn} \citep{gierlinski99} which, differently from the former, assumes zero torque at the inner boundary and the process-dependent parameters are separated from the geometrical parameters, the later making up the norm of this component. The seed photon temperature in \texttt{nthComp} was tied to maximum disk temperature of \texttt{diskpn}. The photon index and electron temperature were tied across \texttt{nthComp} and \texttt{relxillCp}. Similarly, the inner-disk radius was tied across \texttt{diskpn} and \texttt{relxillCp}. The binary inclination for J$1659$ is constrained between 65$^\circ$ - 80$^\circ$ owing to the detection of dips in the light curve and non detection of eclipses \citep{kuulkers13}. However, it is possible for the inner disk to have a different inclination due to the Bardeen-Peterson effect \citep{nealon15}. Thus we relaxed this limit and let the inclination vary between 30$^\circ$ - 85$^\circ$. The best-fit parameters are listed in the third column of table \ref{table:bestfit_par} under Model 3. The fit was excellent ($\chi^2_\nu \sim 1$) but the value of the spin parameter pegged at the negative extreme of $-0.998$ and could not be constrained. An upper limit on R$_{in}$ was found to be $\sim 16$ R$_g$ at 95\% confidence, indicating that the inner disk radius is close to the ISCO. The significance of the \texttt{relxillCp} component was verified by an F-test, the probability of which came out to be $9.49 \times 10^{-10}$. \texttt{diskpn}, being a non-relativistic model, assumes zero spin, and hence, can not be used in a model that measures spin directly. The rationale for using it will be discussed in the next section.

Being a general relativistic disk model, \texttt{kerrbb} is appropriate to characterize the thermal component of the spectrum \citep{li05}. Hence, \texttt{diskpn} was replaced by \texttt{kerrbb} for further analysis. Since the system has a relatively high inclination, the effects of limb darkening were included in the model calculation. The effect of self-irradiation, however, was ignored and a zero torque was assumed at the inner boundary. The spectral hardening factor was fixed at the canonical value of 1.7 \citep{shimura95}. The spin parameter was tied across \texttt{kerrbb} and \texttt{relxillCp} and kept free. This has the advantage of undoing any effect of pile-up that would have remained in the spectrum in-spite of removing the central rows. As concluded by \citet[][]{miller10}, the presence of pile-up in a spectrum would artificially lead to a low spin value upon using reflection spectroscopy and a high spin value upon using continuum fitting method. Hence, tying up the spin from both models would reduce the effect.

After fitting, it was observed that the data cannot constrain all the free parameters, including the spin. In order to freeze the geometrical parameters, prior knowledge on them is required which is derived from the literature. The distance to J1659 is 4.5 - 8.5 kpc \citep{homan13}. This range encompasses the prediction from several other observations \citep{yamaoka12,kennea11}. Similarly, the mass of the black hole is 3 - 10 M$_{\odot}$ \citep{yamaoka12}). As described in the previous section, the inclination was allowed to vary between 30$^\circ$ to 85$^\circ$. Then, a scheme was devised in which the entire parameter space was systematically explored, fixing the geometry parameters to a set of values encompassed within the acceptable range. The grid consisted of the following values: $M=(4, 6, 8, 10)$ $M_{\odot}$ and $D=(4.5, 6.5, 8.5)$ kpc.  After that, the spin was also fixed to a set of 8 equi-spaced values ranging from $-0.998$ to $0.4$. For each of these 216 combinations, the data were fitted for mass accretion rate, $\dot{M}$. Meaningful values of $\dot{M}$ would give us a constrain on the spin.

\begin{deluxetable*}{ccccc}
\tablecolumns{5}
 \tablecaption{Best fit parameters of models as defined in figure \ref{fig:delchi}.}
\tablehead{\colhead{Model components} & \colhead{Parameters} & \colhead{Model 1} & \colhead{Model 2} & \colhead{Model 3}}
 \startdata
 phabs & nH (cm$^{-2}$) & $0.22 \pm 0.01$ & $0.21 \pm 0.01$ & $0.22^{+0.02}_{-0.01}$ \\ 
 gaussian & LineE (keV) & ... & $6.78^\star$ & ... \\
 & Sigma (keV) & ... & $1.54^{+0.50}_{-0.32}$ & ... \\
 & norm ($\times 10^{-2}$) & ... & $1.14^{+0.75}_{-0.36}$  & ...  \\
 diskbb & kT$_{\text{in}}$ (keV) & $0.39 \pm 0.01$ & $0.40^{+0.02}_{-0.01}$ & ...  \\
 & norm ($\times 10^3$) & $7.61^{+1.20}_{-1.01}$ & $5.91^{+1.15}_{-1.04}$ & ...  \\
  diskpn & kT$_{\text{max}}$ (keV) & ... & ... & $0.40^{+0.02}_{-0.01}$ \\
 & R$_{\text{in}}$ (R$_{\text{g}}$) & ... & ... & $10.65^{+5.25}_{}$\\
 & norm ($\times 10^{-2}$)& ... & ... & $9.09^{+0.99}_{-6.53}$\\
 nthComp & $\Gamma$ & $1.87 \pm 0.01$ & $1.93 \pm 0.02$ & $1.90 \pm 0.02$  \\
 & kT$_{\text{e}}$ (keV) & $10.44^{+0.52}_{-0.46}$ & $12.94^{+1.41}_{-1.06}$ & $11.01^{+0.53}_{-0.62}$  \\
 & norm & $0.82 \pm 0.03$ & $0.83 \pm 0.04$ & $0.79^{+0.04}_{-0.02}$\\
 relxillCp & $a$ & ... & ... & $-0.998^\star$  \\
  & Incl (degrees) & ... & ... & $85_{-1.06}$ \\
  &log$\xi$ &...& ... & $2.98^{+0.10}_{-0.24}$  \\
  &A$_{\text{Fe}}$ & ...&... & $10.0_{-1.25}$ \\
  &norm ($\times 10^{-4}$) &...& ... & $3.26^{+1.30}_{-0.08}$ \\
  $\chi^2$ (dof) &... & 244.59 (185) & 139.26 (182) &  173.27 (179) \\
  $\chi^2_\nu$ & ...& 1.32 & 0.76 & 0.97 \\
\enddata
 \tablecomments{The asterisk superscript represents that the parameter pegged to that value while fitting.}
 \label{table:bestfit_par}
\end{deluxetable*}

In the above analysis, the spin was constrained partly through the Fe line, and partly through the continuum. In order to constrain the spin only with the continuum, \texttt{relxillCp} was replaced by a \texttt{Gaussian} to account for the line. Now, it would only be the spin parameter, $a$, in \texttt{kerrbb} that constrains the spin. The black-hole mass and distance were fixed to the grid of values as defined above. Then, the same exercise was repeated by fitting for $\dot{M}$ keeping the spin fixed to a set of values.

The results of the above two exercises are represented in figure \ref{fig:mdotvsa}. Each colored line on the plots represents the combination of distance and black-hole mass that gave a good fit (i.e., $\chi^2_\nu \leq 2$). The four colors denote the four masses chosen. Those combinations of the parameters which did not return a statistically acceptable fit were ignored and are not included in the plots. As is expected, $\dot{M}$ decreases monotonically with increasing spin. This is so, because in \texttt{kerrbb} the inner radius of the accretion disk is assumed to be at the ISCO. So, with increasing spin (i.e., lowering the inner radius), the accretion rate has to decrease to keep the flux constant.

It has been a standard practice to restrict spin measurements to the soft state when the inner accretion disk is presumably at the ISCO. However, it has been shown that the disk extends down to the ISCO even in the hard state if the source is substantially luminous, and robust estimates of that spin have also been given in the hard state \citep{garcia15,miller15}. A theoretical limit of 0.08\% of $\dot{M}_{edd}$ on accretion rate, which translates in to $\sim 0.008$ $L_{edd}$ assuming an efficiency of 0.1, was given by \citet[][]{esin97} below which the disk would be truncated. Similarly, \citet[][]{reynolds13} and \citet{reis10} have studied several XRBs to enunciate observational limits of $0.001$ $L_{edd}$ and $0.0015$ $L_{edd}$ respectively. The luminosity of J1659 lied between 0.019 - 0.067 of the Eddington value for a 10 $M_\odot$ black hole. This range is entirely above both the theoretical and observational limits provided and hence it is possible that no significant truncation of the accretion disk has taken place.

\section{Discussion} \label{sec:discussion}

\begin{figure*}
\gridline{\fig{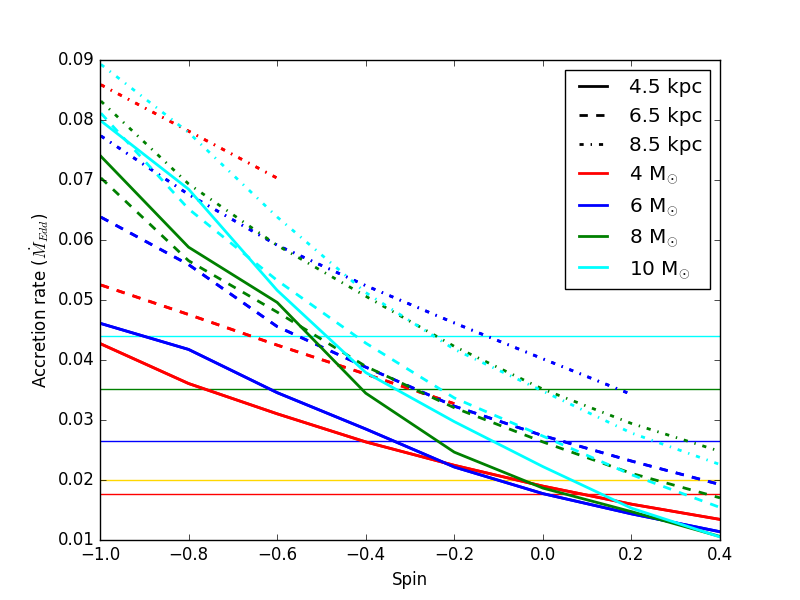}{0.5\textwidth}{(a)}
          \fig{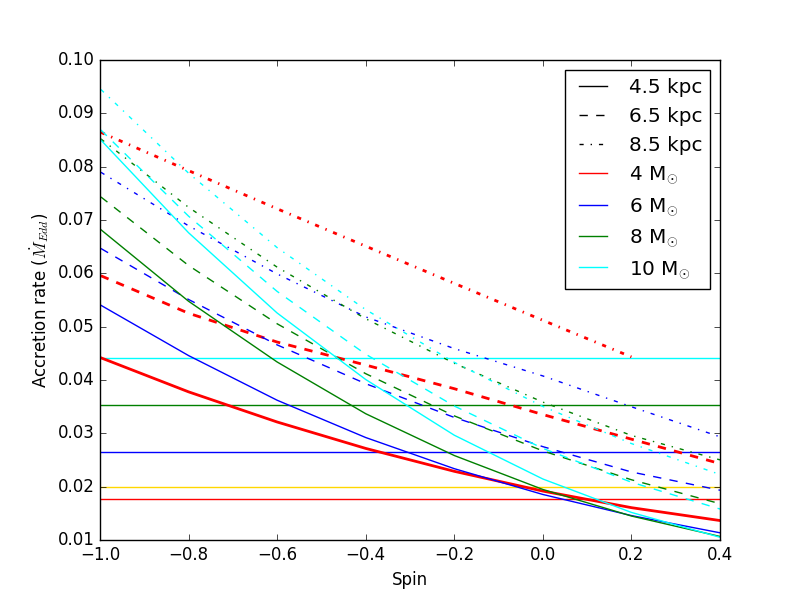}{0.5\textwidth}{(b)}
          }
\caption{Variation of $\dot{M}$ with $a$. Plot (a) represents the results using both CF and Fe line method; plot (b) represents results from only CF method. The different linestyles correspond to the distances and the colors of the lines represent the different black-hole masses. The horizontal lines in the figures represent different lower limits of the accretion rate, as explained in the text.} 
\label{fig:mdotvsa}
\end{figure*}

We carried out a broad-band spectral analysis, using simultaneous XMM-Newton and RXTE data to estimate the spin of the black hole in J1659. We detected a broad Fe line with high significance, which was verified by Monte-Carlo simulation. This allowed us to use reflection spectroscopy along with the continuum fitting method. Due to uncertainties on the geometrical parameters, we employed a novel technique to scan the entire parameter space and represent the accretion rate as a function of spin. Figure \ref{fig:mdotvsa} shows that for reasonable estimates of the mass accretion rate, most of the system parameters unambiguously yield a negative spin. A large fraction of the best-fit parameters also reveal a fascinating and unprecedented consequence of extreme retrograde motion ($a =-1$) for a stellar-mass black hole. These results were ratified by both reflection spectroscopy and continuum fitting method.

Depending on the accretion rate, which is usually quite difficult to ascertain, an upper limit on the spin can be arrived at. We explore different avenues to find a reasonable estimate of the accretion rate, given that a firm lower limit on $\dot{M}$ would be useful in constraining the spin. One such limit can be deduced by considering the fact that the peak luminosity during major outbursts almost always exceeds 8\% of the Eddington limit ($L_{Edd}$) and reaches about 50\% on most occasions \citep{steiner13}. The peak phase of the outburst for J$1659$ lasted for about 25 days (Figure \ref{fig:maxilc}) where the flux hovered between $250$ - $300$ mCrab. The flux at the thermal peak, which occurred on MJD $55489$ is close to that during our observation with a flux of $\approx 260$ mCrab \citep{kalamkar11}. Since luminosity $L = F\times4\pi D^2 \propto \dot{M}$ for radiatively efficient accretion \citep{frank02}, the accretion rate during our observation should be comparable to that during the peak. A lower limit of $8\%$ on peak accretion rate constrains the spin to extreme negative values. The plateau phase of J$1659$ was also associated with a few flaring events which were, however, not associated with changes in the spectral hardness \citep{kalamkar11}. These flares pose an ambiguity in the choice of the outburst-peak. Nevertheless, even considering the strongest flare on MJD $55477$ to represent the peak, the flux during our observation is only a factor of $\sim 1.5$ lower than at this peak. This leads to the accretion rate being $\sim 5.3 \%$ of $\dot{M}_{Edd}$ during our observation which also entirely restricts the spin to negative values.  \\
Another limit comes from the norm of \texttt{diskpn} from the fit using Model 3. Using the formalism laid out by \citet[Appendix A,][]{gierlinski99}, the accretion rate can be expressed as a function of black-hole mass, maximum disk temperature, and inner-disk radius. The different $\dot{M}$ values calculated using the best-fit values of the above parameters are represented in the figure \ref{fig:mdotvsa} through horizontal lines, the color of which corresponds to each black-hole mass chosen. Although \texttt{diskpn} is a non-GR model, assuming a static black hole, the accretion rates obtained from fits with this component are consistent with the ones from \texttt{kerrbb} having a significant overlap in the parameter space. This overlapping region also falls almost entirely in the negative spin domain, with an upper limit of $\sim 0.2$ for Fe-line method and $\sim 0.4$ for CF method. The fact that our fits favor a negative spin implies that the inner disk radius remains farther away than 6 R$_g$, thus justifying the use of \texttt{diskpn}.\\
To be fastidious enough, a much firmer limit on the black-hole spin in J1659 can be obtained by considering the fact that for the thin accretion disk to exist, the accretion rate has to be at least 2\% of the Eddington limit \citep{narayan98,meyer00}. Below this limit, the accretion flow would be in the form of an ADAF, with the X-ray luminosity being too low. An $\dot{M}$ of 0.02, in Eddington units, gives a higher and a more conservative upper limit, a prograde but moderately rotating black hole. Our analysis of J$1659$ \citep[see also][]{kalamkar11,yamaoka12}, shows that this limit is most likely an overkill since a thermal disk component with a modest temperature of about 0.4 keV is indispensable for fitting the data.

Finally, we also test the possibility of a truncated prograde disk at the expense of other parameters. The spin was fixed to three values of 0, 0.3 and 0.9 while keeping the black-hole mass, distance and inclination unconstrained and free to vary. Since the geometrical parameters were left free, the statistics remained reasonably good and did not change drastically as in the earlier case ($\triangle \chi^2 \approx 13$ per d.o.f). Best-fit value of black-hole mass and inclination were slightly higher, but acceptable. However, the best-fit value of accretion rate attained much lower values of $0.18\%,0.09\%$ \& $0.004\% $ of $\dot{M}_{Edd}$ respectively. These values are too low, even for the formation of the thin accretion disk \citep{narayan98}. The distance and ionization parameters were also constrained to unphysically lower values. This shows that the data preferred a truncated prograde disk only for unphysical values of accretion rate, distance and ionization parameter.  
With this we demonstrate an unambiguous detection of retrograde spin for a stellar-mass black hole which is independent of the choice of the black-hole geometric parameters, and is concurrent across both Fe-line spectroscopy and continuum fitting method. This result further opens up the possibility that retrograde motion among black holes is a norm rather than exception.

\bibliography{sample63}{}
\bibliographystyle{aasjournal}

\end{document}